\documentstyle[preprint,aps,epsfig]{revtex}
\tighten
\def\bea{\begin{eqnarray}}
\def\beann{\begin{eqnarray*}}
\def\beq{\begin{equation}}
\def\eea{\end{eqnarray}}
\def\eeann{\end{eqnarray*}}
\def\eeq{\end{equation}}
\def\nn{\nonumber}

\newcommand{\bcdot}{\bbox{\cdot}}

\newcommand{\bsigma}{\bbox{\sigma}}

\newcommand{\btau}{\bbox{\tau}}

\begin{document}
\headheight1.2cm
\headsep1.2cm
\baselineskip=20pt plus 1pt minus 1pt
%\tolerance=1500
%
%
\preprint{
\vbox{
\hbox{IFT-P.014/99, ADP-99-10/T355}
}}
%%%%%%%%%%%%%%%%%%%%%%%%%%%%%
\draft
\title{Chiral corrections in hadron spectroscopy}
\author{
A.W. Thomas$^{1}$ and  G. Krein$^2$ \\
{\small $^1$ Department of Physics and Mathematical Physics and Special 
Research Center for} \\
{\small the Subatomic Structure of Matter, University of Adelaide, SA 5005, 
Australia} \\
{\small  $^2$ Instituto de F\'{\i}sica Te\'{o}rica, Universidade Estadual 
Paulista}\\
{\small Rua Pamplona, 145 - 01405-900 S\~{a}o Paulo, SP, Brazil}\\
}
\maketitle
\begin{abstract}
We show that the implementation of chiral symmetry in recent studies of the
hadron spectrum in the context of the constituent quark model is inconsistent
with chiral perturbation theory. In particular, we show that the leading 
nonanalytic (LNA) contributions to the hadron masses are incorrect in such 
approaches. The failure to implement the correct chiral behaviour of QCD 
results in incorrect systematics for the corrections to the masses.
\end{abstract}
\vspace{2.5cm}
\noindent{PACS NUMBERS: 24.85.+p, 11.30.Rd, 12.39.Jh, 12.39.Fe, 12.40.Yx}

\vspace{1.0cm}

\newpage 

There is an extremely interesting recent series of papers by Glozman, Riska 
and collaborators~\cite{GR1}-~\cite{SPG} who have investigated hadron
spectroscopy on the basis of a residual $q-q$ interaction governed by
chiral symmetry. Their residual interaction, which is meant to correspond to
Goldstone boson (GB) exchange, has the attractive feature, in comparison
with one-gluon-exchange (OGE)~\cite{deRGG}, that it does not produce large
spin-orbit effects which are certainly not present in the spectrum. While
our remarks apply to all GB exchanges, for simplicity we concentrate on
the SU(2) sector -- i.e. pion exchange. In this sector GB exchange leads to 
an effective interaction of the form 

\beq
H_{int} = \frac{g^2}{4\pi} \frac{1}{3} \sum_{i < j} \bsigma_i\bcdot\bsigma_j
\btau_i\bcdot\btau_j \left[ m^2_{\pi} \frac{ e^{-m_{\pi} r_{ij}} }{r_{ij}}
- 4\pi \delta(r_{ij})\right],
\label{OPE-spin}
\eeq

\noindent
where $m_i$ and $m_j$ denote the masses of the constituent quarks and $m_\pi$
is the pion mass. In principle there is also a tensor component, which will
not be written explicitly since it is not relevant in the context of the 
present paper. This interaction has also been employed in studies of the
hadron properties and hadron-hadron 
interactions~\cite{{NN-rev},{recent}}. In practice, the short-distance 
behaviour of this interaction is not expected 
to be reliable~\cite{GR1} - unlike the long range Yukawa piece - and in the 
spectroscopic studies by Glozman and Riska the radial strength is replaced 
by single fitting parameter in each shell. On the other hand, the 
spin-isospin structure of Eq.~(\ref{OPE-spin}) is maintained and the 
corrections from Eq.~(\ref{OPE-spin}) to the energy of the nucleon (N) and 
the $\Delta(1232)$ are given as

\bea
M_N &=& M_0 - 15 P^{\pi}_{00} \label{delMN} \\
M_{\Delta} &=& M_0 - 3 P^{\pi}_{00} \label{delMD},
\eea

\noindent
where $M_0$ is the corresponding unperturbed energy and $P^{\pi}_{00}$ is 
the fitting parameter corresponding to the radial matrix element of 
Eq.~(\ref{OPE-spin}), in the lowest-energy unperturbed shell of the 3-quark 
system. 

Because the basis for this approach to hadron spectroscopy is chiral 
symmetry, we were interested to check that the formalism is consistent with 
chiral perturbation theory ($\chi$PT) - i.e., that at least the leading 
nonanalytic (LNA) contribution to hadron masses is correct. It turns out to 
be very easy to check this and the result is that Eq.~(\ref{OPE-spin}) is 
inconsistent with the LNA behaviour of QCD.

The LNA contribution to the mass of the nucleon is proportional to 
$m^3_{\pi} \sim m^{3/2}_q$~\cite{LP}. In the quark model of Glozman and 
Riska, such a contribution can only arise from the linear term in the 
expansion of the Yukawa potential in Eq.~(\ref{OPE-spin})

\bea
H^{LNA}_{int} &=& \frac{g^2}{4\pi} \frac{1}{3} \sum_{i < j} 
\bsigma_i\bcdot\bsigma_j \btau_i\bcdot\btau_j m^2_{\pi} 
\frac{{1-m_{\pi}r_{ij}+{\cal O}(m^2_{\pi})}}{r_{ij}}\nn\\
&\sim & - m^3_{\pi} \frac{g^2}{4\pi} \frac{1}{3} 
\sum_{i < j} \bsigma_i\bcdot\bsigma_j \btau_i\bcdot\btau_j .
\label{LNA-OPE}
\eea
 
\noindent
The radial matrix element is therefore a normalization integral and hence 
model independent, as it must be. The overall strength (in hadron 
$|H\rangle$) is given by the spin-isospin matrix element

\beq
\langle SI \rangle_{H} = \langle H |\sum_{i < j} \bsigma_i\bcdot\bsigma_j 
\btau_i\bcdot\btau_j |H\rangle .
\label{SI}
\eeq

\noindent 
For the N and the $\Delta$ this gives

\bea
\langle SI \rangle^{\text{Eq.}\;(1)}_{N} &=& 30 \label{GRLNAN} \\
\langle SI \rangle^{\text{Eq.}\;(1)}_{\Delta} &=& 6 . \label{GRLNAD} 
\eea

On the other hand, the corresponding matrix elements from the LNA 
contribution required by $\chi$PT are given by~\cite{J}

\bea
\langle SI \rangle^{\chi}_{N} &=& 25 \label{QCDLNAN} \\
\langle SI \rangle^{\chi}_{\Delta} &=& 25. \label{QCDLNAD}
\eea

\noindent
The formulation of $\chi$PT including the $\Delta(1232)$ as an explicit
degree of freedom was originally proposed in Ref.~\cite{JM}.
These contributions arise from the processes shown in Figs.~(1a) and (b), 
respectively. We stress that this requires, as usually assumed in 
$\chi$PT, that the $N$ and $\Delta$ are not degenerate in the chiral limit. 
For a critical discussion on this subject, we refer the reader to 
Ref.~\cite{BM}. The LNA chiral contributions to the octet and decuplet 
baryons has also been calculated within the framework which combines the
1/$N_c$ expansion with $\chi$PT, where $N_c$ is the number of colors. 
Large $N_c$ $\chi$PT was originally proposed by Dashen and Manohar~\cite{DM},
and has been further developed by many authors (for a list of references, see
Ref.~\cite{OW}).

\vspace{1.0cm}
\begin{center}
\epsfig{angle=0,figure=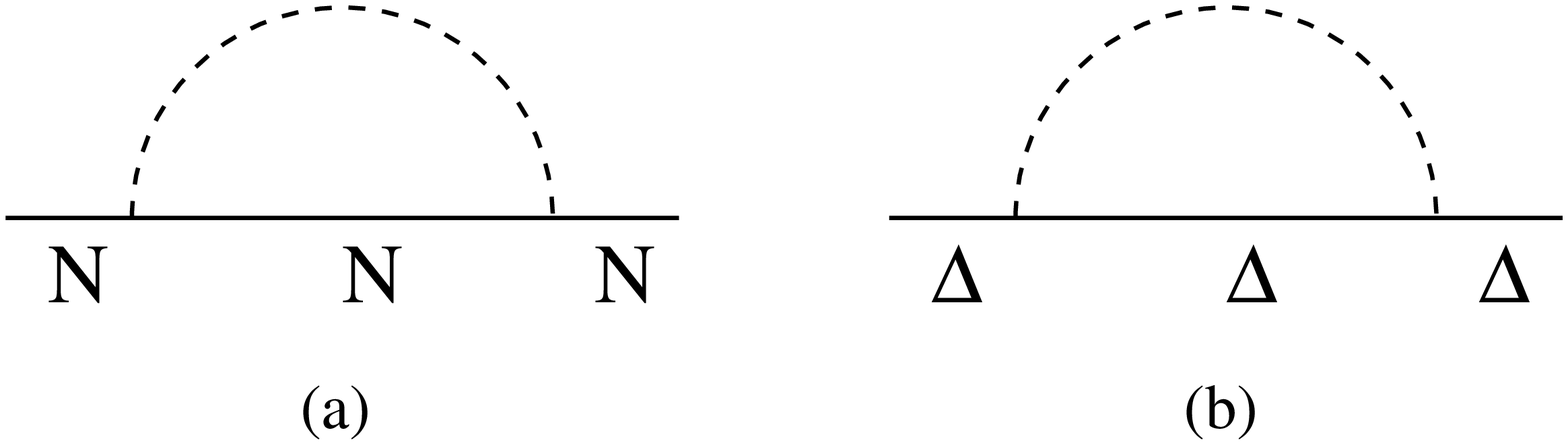, height=3.0cm}
 
\vspace{0.5cm}FIGURE 1. One-loop pion self-energy of (a) the nucleon (N) and 
(b) the delta ($\Delta$).
\end{center}

\vspace{1.0cm}
A comparison of these results shows that Eq.~(\ref{OPE-spin}) yields the
wrong LNA contribution for both the $N$ and the $\Delta$. For the N, the 
error is not large ($30$ compared to $25$). However, because the error is 
much larger for the $\Delta$ the crucial point is that the {\em systematics} 
are {\em wrong}. For example, with the correct coefficients this mechanism 
provides no $\Delta$-N mass difference at all! Of course, our arguments 
concern the systematics of the LNA behaviour implied by Eq.~(\ref{OPE-spin}).
Even though the Yukawa term is not actually used in the spectral studies, the
coefficient of the short-range piece, which {\em is} used, is the same and
hence our arguments are directly relevant to the actual calculations.

In $\chi$PT the LNA contribution to the nucleon mass is given by

\beq
M^{LNA}_{N} = - \frac{3}{32 \pi f^2_{\pi}} g^2_A m^3_{\pi},  
\label{LNAChPT}
\eeq

\noindent
where $f_\pi \sim 93$~MeV is the pion decay constant and $g_A =$~1.26 is
the weak decay constant. In a quark model, the crucial step in ensuing this 
LNA behaviour is to project the quark states onto bare baryon 
states~\cite{CBM}. Specifically, in a constituent quark model of the
Glozman-Riska type, the bare states would correspond to the three quark
states confined by a phenomenolical potential. The effective {\em hadronic} 
Hamiltonian is obtained by projecting the quark-model Hamiltonian, which now
includes the quark-pion vertices, on the basis of the bare three-quark states.
Chiral corrections to hadronic properties, such as masses and magnetic 
moments, are then calculated in time-ordered perturbation theory with the
effective hadronic Hamiltonian. For a constituent quark model of the 
Glozman-Riska type, such a procedure leads to corrections to the N and the 
$\Delta$ masses of the form

\bea
M_N &=& M^{(0)}_N - \frac{3}{16\pi^2 f^2_{\pi}} g^2_A \int_0^{\infty} dk\,
\frac{k^4\,u^2_{NN}(k)}{w^2(k)} - \frac{3}{16 \pi^2 f^2_{\pi}} \,
\frac{32}{25} g^2_A \int_0^{\infty} dk\, \frac{k^4\,u^2_{N\Delta}(k)}{w(k)\left(\Delta M+w(k)
\right)} \label{MN} \\
M_\Delta &=& M^{(0)}_{\Delta} + \frac{3}{16\pi^2 f^2_{\pi}} \,\frac{8}{25} 
g^2_A \int_0^{\infty} dk\, \frac{k^4\,u^2_{N\Delta}(k)}
{w(k)\left(\Delta M-w(k)\right)} - \frac{3}{16\pi^2 f^2_{\pi}} \,g^2_A 
\int_0^{\infty} dk\, \frac{k^4\,u^2_{\Delta \Delta}(k)}{w^2(k)} .
\label{MD}
\eea

\noindent
Here, the $M^{(0)}$'s are the masses in the chiral limit, 
$\Delta M = M_{\Delta} - M_N$, $g_A~=~$5/3 is the bare axial coupling given 
by the constituent quark model, $w(k)=\sqrt{k^2+m^2_{\pi}}$ is the pion 
energy and $u_{NN}(k)$, $u_{N\Delta}(k)$, $\dots$ are the $NN\pi$, 
$N\Delta \pi$, $\dots$  form factors. The LNA contribution to $M_N$ is 
easily seen to arise from the first integral in Eq.~(\ref{MN}) 
(c.f. Fig.~1(a)), while the LNA contribution to $M_{\Delta}$ comes from the 
second integral in Eq.~(\ref{MD}) - c.f. Fig.~1(b).

In order to understand why the use of Eq.~(\ref{OPE-spin}) is wrong, we 
consider the limit, generally considered physically 
unlikely~\cite{JM}\cite{BM}, that $\Delta M = 0$. Then all integrals in 
Eqs.~(\ref{MN}) and (\ref{MD}) have the same LNA behaviour and the 
contributions are in the ratio

\bea
25\;(N \rightarrow N\pi \rightarrow N)\;\;&:&\;\;32\;(N \rightarrow 
\Delta\pi \rightarrow N) \label{NDN}\\
8\;(\Delta \rightarrow N\pi \rightarrow \Delta)\;\;&:&\;\;25\;
(\Delta \rightarrow \Delta\pi \rightarrow \Delta). \label{DND}
\eea

\noindent
In this case the ratio of the total $N$ and $\Delta$ self-energies is 
$57~:~33$ and the difference is identical to that given by 
Eqs.~(\ref{GRLNAN}) and (\ref{GRLNAD}). This 
recalls the well known result from the early work on chiral bag 
models~\cite{{Jaffe},{CTR},{MBX},{CBM}} that the calculation of the 
self-energy integrals through projection on all baryon states in which 
the orbital quantum numbers are unchanged, in the limit where these are
degenerate, is equivalent to calculating pion emission and absorption 
between {\em all} quarks. In particular one must include those diagrams 
where the pion is emitted and absorbed by the same quark. In this case the 
spin-isospin structure of the pion interaction~is

\beq
\langle SI \rangle_H = \frac{1}{2} \sum_{ij} \langle H |
\bsigma_i\bcdot\bsigma_j\,\btau_i\bcdot\btau_j |H\rangle ,
\label{i=j}
\eeq

\noindent
so that $\langle SI \rangle_N=57$ and  $\langle SI \rangle_\Delta=33$ --
which agree with the results based on Eqs.~(\ref{MN}) and (\ref{MD}),
quoted in Eqs.~(\ref{NDN}) and (\ref{DND}).
Precisely this form of the pion self-energy was suggested in the early 
spectroscopic study of Mulders and Thomas~\cite{MT} -- see also 
Refs.~\cite{{CTR},{MBX},{PLF}}, and Ref.~\cite{GhP} for more recent work.

For completeness, we remark that the ratios given in Eqs.~(\ref{NDN}) and
(\ref{DND}) are precisely the leading order corrections given by large $N_c$ 
$\chi$PT, as can be easily checked making use of Eqs.~(5.6), (C1) and C(5) of 
Ref.~\cite{OW}. 

In practice, the $\Delta-N$ mass difference is quite large and the 
contribution from the process $N \rightarrow \Delta \pi \rightarrow N$ is 
consequently suppressed. One would still expect to obtain a sizeable 
fraction of the $\Delta-N$ splitting from pion exchange. Indeed, at the 
price of increasing the size of the pion-quark effective coupling one could 
refit the whole mass difference in terms of pion exchange. This would have 
the consequence that the total nucleon self-energy associated with pion 
exchange would need to be of the order of $700$~MeV.  Whether one is able 
to live with such large self-energies remains to be seen. The alternative 
is to add some additional hyperfine interaction, such as gluon exchange or 
residual instanton effects.

In conclusion, we repeat that the use of Goldstone boson exchange
interactions of the type given in Eq.(1) is inconsistent with the chiral
structure of QCD. In order to reproduce the correct chiral behaviour 
one must include Goldstone boson exchange between {\em all} quarks, 
including self-interactions, but the intermediate quark states must 
be projected onto (bare) baryon states -- as carried out, for example, 
within the Cloudy Bag Model~\cite{CBM}. While our analysis of the 
spectroscopic studies of Glozman and collaborators shows that these are 
incomplete, the {}findings are not entirely negative. One can still hope that 
the major qualitative features of this work will survive in a complete 
re-analysis. Such a re-analysis must now be an urgent priority. 

\vspace{1.0cm}
This work was supported by the Australian Research Council and CNPq (Brazil).
One of us (AWT) would like to acknowledge the warm hospitality of the
Institute for Theoretical Physics at UNESP, where much of the work was 
carried out.

\end{document}